\documentclass[reqno,11pt,intlimits]{amsproc}
\usepackage{fixltx2e}
\usepackage{amssymb}
    \numberwithin{equation}{section}

\usepackage[nobysame,citation-order]{amsrefs}
\usepackage{xyzbib}

\usepackage[foot]{amsaddr}

\usepackage{hyperref}

\newcommand\lmove{\hspace{-.25in}}

\makeatletter
\def\@setemails{%
   \lmove\mbox{{\itshape E-mails}:\space}{\ttfamily\emails}.
 }
\makeatother

\newcommand\rme{\mathrm{e}}

\renewcommand\leq\leqslant
\renewcommand\geq\geqslant

\newcommand\Dex{\Lambda}
\newcommand\Hex{\hat H}
\newcommand\Path{\mathcal{P}}
\newcommand\Up{\Uparrow}

\newcommand\bra[1]{\langle\, #1\, \rvert\,{}}
\newcommand\ket[1]{{}\,\lvert\, #1\, \rangle}
\newcommand\bracket[2]{\langle\, #1\,\mid\, #1\, \rangle}

\begin{document}
\vspace*{0in}

\title{Scaling of many-particle correlations in a dissipative sandpile}

\author{N. M. Bogoliubov$^{1)}$}
\email{bogoliub@pdmi.ras.ru}

\author{A. G. Pronko$^{1,2)}$}
\email{agp@pdmi.ras.ru}

\author{J. Timonen$^{3)}$}
\email{jussi.t.timonen@jyu.fi}

\address{\lmove$^{1)}$Saint Petersburg Department of V.A. Steklov Mathematical Institute of
Russian Academy of Sciences, Fontanka 27, 191023 Saint Petersburg, Russia}
\address{\lmove$^{2)}$Department of Physics, University of Wuppertal, 42097 Wuppertal, Germany}
\address{\lmove$^{3)}$Department of Physics, University of Jyv\"askyl\"a, P.O. Box 35 (YFL),
40014 Jyv\"askyl\"a, Finland}


\begin{abstract}
The two dimensional directed sandpile with dissipation is
transformed into a $(1+1)$ dimensional problem with discrete space and
continuous `time'. The master equation for the conditional probability that
$K$ grains preserve their initial order during an avalanche can thereby be
solved exactly, and an explicit expression is given for the asymptotic form
of the solution for an infinite as well as for a semi-infinite lattice in the
horizontal direction. Non-trivial scaling is found in both cases. This
conditional probability of the sandpile model is shown to be equal to a
$K$-spin correlation function of the Heisenberg XX spin chain, and the
sandpile problem is also shown to be equivalent to the `random-turns' version
of vicious walkers.
\end{abstract}

\maketitle

\section{Introduction}
\label{intro}

Non-equilibrium dynamic systems have been for some time of considerable
interest as they can exhibit critical behaviour in close analogy with
systems at thermal equilibrium. A certain class of such dynamic systems,
various sandpile models \cite{DR-89,D-90,DP-04,M-91,KMS-05,TK-00,SCh-05}, have
become a standard framework when analyzing self-organized criticality
\cite{BTW-88,A-03}, \textit{i.e.} when the dynamics of the system
inevitably drives it to a critical state independent of the initial
state. Despite the extensive work on these systems, it is only fairly
recently that a more detailed understanding of problems like when exactly
sandpile models exhibit self-organized criticality, or what are the
possible universality classes of their critical behaviours, have begun to
emerge.

Most of the work so far on sandpiles has thus concentrated on
properties such as \textit{e.g.} the average duration of avalanches and
their size distribution, which both exhibit scaling in a critical state.
However, there may well be for example interesting many-particle
correlations in sandpiles, which likewise exhibit scaling. If one
considers particles with non-intersecting trajectories, interesting
connections with problems like vicious walkers \cite{F-84,HF-84} would
probably arise. Non-intersecting Brownian walkers have also been of very
recent interest, and they as well seem to display corresponding scaling
properties \cite{FMS-10}.

It is the `fermionic' nature of vicious (and non-intersec\-ting Brownian)
walkers, which gives rise to scaling of their (asymptotic) survival
probability, and the related scaling exponent depends in a nontrivial way
on the number of walkers, as well as on the boundary conditions imposed
\cite{Fo-91,Fo-01,KT-02,KTNK-03,SMCR-08,ShB-03}.
One would thus expect that suitably defined conditional (many-particle)
probabilities of particles with non-intersecting paths in sandpiles should
exhibit rather similar properties. If the trajectories of vicious walkers
and non-intersecting Brownian particles are equivalent to `worldlines' of
free fermions, one would expect in addition that these sandpile
probabilities can be transformed into a problem of free fermions, or,
equivalently, into one of spin ($S=1/2$) chains.

In order to address these questions, we consider in this paper the
probability that $K$ particles of a two dimensional (2D) directed
sandpile of ref. \cite{DR-89} preserve their initial order during
an avalanche. First we reformulate the (Abelian) sandpile model such
that it becomes one in 1+1 dimensions, and assume for the sake of
generality that it is dissipative, \textit{i.e.} that the number of
`grains' is not conserved in the topplings of unstable sites. We also
consider two different boundary conditions, an infinite system (in the
horizontal direction) and a system with an absorbing boundary at the
origin (a `semi-infinite' system). As it is well known by now, the model
is critical only at vanishing dissipation \cite{TK-00,VD-01}. We derive
an exact analytic form for the probability, and show that non-zero
dissipation introduces an exponential cutoff in its asymptotic form
that also includes a power law with a scaling exponent that depends
nonlinearly on $K$, and is different for the two boundary
conditions.

We also show that this probability is equal to the partition
function of $K$ vicious walkers, more precisely the `random-turns'
version of such walkers \cite{F-84,Fo-01}. The former probability is
thus the generating function for the survival probabilities of the
walkers. The scaling exponents of the sandpile probability are not
those of the survival probability of the `lock-step' vicious walkers,
although the two problems are intimately connected. Finally we show
that the sandpile probability is equal to a correlation function of
the Heisenberg XX spin chain. This establishes the relation of
both the sandpile problem and the walker problem to a problem of
free fermions, as the Heisenberg XX chain is equivalent to free fermions via
a Jordan-Wigner transformation.

\section{Abelian sandpile model}
\subsection{Discrete model}

A 2D directed (Abelian) sandpile model on a lattice (see, e.g., \cite{D-90})
is constructed such that to each site $(j,n)$ an integer height variable
(number of grains) $z_{(j,n)}$ is assigned. The site has a threshold
height $z_{(j,n)}^c$ below which it is stable. The dynamics of the
model consists of two steps. First, we choose a site $(j,n)$ at random
and add one grain to it, \textit{i.e.}, $z_{(j,n)}\mapsto z_{(j,n)}+1$.
For $z_{(j,n)}\geq z_{(j,n)}^c$, site $(j,n)$ becomes unstable and its grains
are distributed among the 'downhill' neighbouring sites. In the following
we will use the notation by which the locations of lattice sites
in the horizontal direction are labelled by $j$, $k$ or $l$, and
by $n$ in the downhill direction. By $n$ we can equivalently denote the
number of steps in a cascade of toppling processes. In a toppling at site
$(j,n)$ grains are thus distributed to sites $(j+1,n+1)$ and $(j-1,n+1)$.
By supressing the $n$ labels (understanding that two adjacent columns
in the lattice are connected in a toppling and that there is no $n$ dependence)
we can express a toppling in the form
\begin{equation}\label{z}
z_j\mapsto z_j-\Delta _{lj},
\end{equation}
in which the elements of the toppling matrix $\Delta$ satisfy
$\Delta_{jj}>0$, and $\Delta_{lj}<0$ for $l\neq j$. The condition
$\sum_j\Delta_{lj}\geq 0$ for every $l$ guarantees that no grains are
created in the toppling process. Without loss of generality we can put
$\Delta_{jj}=z_{(j,n)}^c$. The allowed number of grains in a stable site
$(j,n)$ is now $1,2,\dots,\Delta_{jj}-1$. The sites $(j,n)$ such
that $\sum_j\Delta_{lj}>0$ are called dissipative. Boundary sites
are always dissipative so that grains can leave the system through
the boundaries. After an initial toppling at a site, neighbouring
sites can also become unstable, and sites are kept on relaxing
with parallel updating until all sites are stable. In this way an
avalanche of topplings is generated. Existence of dissipative
sites ensures that all avalanches terminate in a finite time.

Assume now that all lattice sites are initially in a stationary state
(i.e. are stable): $z_{(j,n)}=z_{(j,n)}^c-1$. If we add a grain at a
randomly chosen site $(l,0)$, and make site $(j,n)$ dissipative such
that the system returns to a stationary state after the extra grain
disappears from this site (i.e. after $n$ steps). The conditional
probability $G_{jl}(n)$ that an extra grain is at site $(j,n)$ satisfies
the equation
\begin{equation} \label{de}
G_{jl}(n)=\frac 12\left\{ G_{j+1l}(n-1)+G_{j-1l}(n-1)\right\},
\end{equation}
with the initial condition $G_{jl}(0)=\delta _{jl}$. Since we
consider only symmetric topplings, the conditional probability also
satisfies $G_{jl}(n)=G_{lj}(n)$. It is easy to verify that
eq. \eqref{de} is the same as the equation for the corresponding
probability expressed in the conventional 'light cone' coordinates,
eq. (5) in ref. \cite{DR-89}.

\subsection{Continuous 'time' model}
\label{sec:continuous}

We can also express eq. \eqref{de} in the form
\begin{equation} \label{de2}
G_{j,l}(n+1)-G_{j,l}(n)=\frac{1}{2}
\left\{G_{j+1,l}(n)+G_{j-1,l}(n)-2G_{j,l}(n)\right\} .
\end{equation}
Consider now a process in which the discrete number of steps is replaced
by a continuous parameter that will be called 'time' in the following.
Let $P_{jl}(t)$ be the
conditional probability that a grain is at a horizontal location $j$ at time $t$
after an arbitrary number of steps since it was dropped at a horizontal location
$l$ at $t=0$. Transforming
eq. \eqref{de2} into such a continuous time we obtain that, during a short time
interval $dt$, the probability $P_{jl}(t)$ changes such that
\begin{equation}
P_{j,l}(t+dt)-P_{j,l}(t)=\frac{1}{2}
\left\{P_{j+1,l}(t)+P_{j-1,l}(t)-2P_{j,l}(t)\right\}\, dt,
\end{equation}
which leads to the master equation
\begin{equation}\label{master}
\frac d{dt}P_{jl}(t)=-\frac 12\sum_k\Delta _{jk}P_{kl}(t)
\end{equation}
with the toppling matrix
\begin{equation}\label{con}
\Delta_{jk}=2\delta_{jk}-(\delta_{j+1,k}+\delta_{j-1,k}).
\end{equation}
This toppling matrix means that, as above, at each toppling two grains
are removed from the site and distributed to its nearest-neighbour
downhill sites. We consider here only symmetric topplings,
$\Delta_{jk}=\Delta_{kj}$, and thus $P_{jk}(t)=P_{kj}(t)$. The
initial conditions are $P_{jk}(0)=\delta_{jk}$.  For a model of
$N$ sites in the horizontal direction, the lateral boundary elements
of the toppling matrix can be defined such that
$\Delta_{0,1}=\Delta_{N,N+1}=0$, and hence the boundary sites $j=1$
and $j=N$ are always dissipative as required.

Notice that the continuous time is not a simple continuum formed by
the discrete variables $n$, but $P_{jl}(t)$ includes processes with
all possible numbers of steps. In fact
function $P_{jl}(t)$ can be considered as the generating
function of the conditional probabilities $G_{jl}(n)$ as we find that
\begin{equation}\label{gf}
e^tP_{jl}(t)=\sum_{n=0}^\infty G_{jl}(n) \frac{t^n}{n!}.
\end{equation}

The expected number of topplings at site $j$ in an avalanche resulting
from a perturbation (adding a grain) at site $l$ is given by
$\Gamma_{jl}(0)=\sum_{n=0}^\infty G_{jl}(n)$. A Laplace transform of
the conditional probability
\begin{equation}
\Gamma_{jl}(f)=\int_0^\infty \rme^{-tf}P_{jl}(t)\, dt,
\end{equation}
is the Green's function of the master equation eq. \eqref{master}.
It is easy to verify that $\Gamma _{jl}(0)$ satisfies \cite{D-90}
the condition $\sum_{k}\Delta_{jk}\Gamma_{kl}(0)=\delta_{jl}$.

The master equation eq. \eqref{master} can easily be solved for the
toppling matrix of eq. \eqref{con} with the initial condition
$P_{jl}(0)=\delta _{jl}$. Consider first the case of an infinite lattice
in the horizontal direction such
that $-\infty < j,l < \infty$. In this case we find that
\begin{equation}\label{infty}
P_{jl}^{(-\infty,\infty)}(t)=\rme^{-t}I_{l-j}(t),
\end{equation}
where $I_{j}(x)$ is a modified Bessel function.
Asymptotically, for large $t$,
the conditional probability for a single grain thus behaves as
\begin{equation}\label{inflat}
P_{jl}^{(-\infty,\infty)}(t)\propto t^{-1/2}.
\end{equation}
There is a pure power law so that the duration of avalanches scales
with an exponent $\xi_{(-\infty,\infty)} =1/2$. This exponent coincides with
the known result for 2D directed
sandpiles \cite{DR-89} as it should.

We can analyze the effect of boundary conditions by introducing
an absorbing boundary at the origin. To this end we first
recall the solution for a finite lattice
of $N$ sites (see \textit{e.g.} \cite{VD-01}) for which the boundary
conditions are $P_{jl}(t)=0$ for $j,l=0$ and $j,l=N+1$. In this case one has
\begin{equation}\label{hc}
P_{jl}(t)=\frac{2}{N+1}\sum_{k=1}^N\rme^{-tE_k}
\sin\frac{\pi j k}{N+1}\sin\frac{\pi l k}{N+1},
\end{equation}
where the spectrum is of the Bloch form,
\begin{equation}\label{Ek}
E_k=1-\cos\frac{\pi k}{N+1}.
\end{equation}

In the limit $N\to\infty$ the sum in eq. \eqref{hc} can be replaced by an integral,
with the result
\begin{align}\label{erinf}
P_{jl}^{(0,\infty)}(t)
&=\frac 2\pi \int_0^\pi e^{-t(1-\cos x)}\sin (lx)\sin (jx)\,dx
\notag\\
&=\rme^{-t}\big[I_{l-j}(t)-I_{l+j}(t)\big].
\end{align}
The asymptotic behaviour for large $t$ of the conditional
probability is now given by
\begin{equation}\label{scl}
P_{jl}^{(0,\infty)}(t)\propto t^{-3/2}.
\end{equation}
The scaling exponent indeed depends on having a boundary at a
finite distance: $\xi_{(0,\infty)} =\frac{3}{2} =\xi_{(-\infty,\infty)}+1$. As expected,
the same exponent has been found for the scaling of avalanche sizes
with the corresponding boundary conditions \cite{TNURP-92,Th-93}.

\section{Multiple-grain correlations}
\subsection{Master equation}

Having established that our master equation method indeed
reproduces previously known results,
we turn now to a more
interesting problem of correlations between multiple grains during
`avalanche dynamics'.
To this end, let us address the following
problem. Consider the same lattice as above with all its
sites in a stationary state: $z_{(j,n)}=z^c_{(j,n)}-1$, and add  $K$
grains at randomly chosen $K$ horizontal locations:
$l_1>l_2>\dots>l_K$. The toppling rules are the same as above, at each toppling two grains are
removed from the toppling site $j$, and a grain can jump to each of the two
nearest-neighbour sites in the downhill direction. However, if
$z_{(j,n)}-z_{(j\pm 1,n+1)}=0$, site $(j,n)$ cannot
topple. The probability that the additional grains will be at dissipative
sites $j_1>j_2>\dots>j_K$ at time $t$ (after an arbitrary number of topplings)
satisfies a generalized version of eq. \eqref{master}, namely
\begin{multline}\label{mpmeq}
\frac{d}{dt} P_{j_1,\dots,j_K;l_1,\dots,l_K}(t)
=\frac 12\sum_{r=1}^K\big[P_{j_1,\dots,j_K;l_1,\dots,l_{r-1},l_r+1,l_{r+1},\dots,l_K}(t)
\\
+P_{j_1,\dots,j_K,l_1,\dots,l_{r-1},l_r-1,l_{r+1},\dots,l_K}(t)\big]
-KP_{j_1, \dots, j_K; l_1, \dots, l_K}(t),
\end{multline}
supplemented by the condition $P_{j_1,\dots,j_K;l_1,\dots,l_K}(t)=0$,
if $j_r=j_{r+1}$, for all $r=1,\dots, K-1$. The solution
to this equation is given by
\begin{equation}\label{det}
P_{j_1\dots j_K,l_1\dots l_K}(t)=
\det_{1\leq r,s \leq K} \{ P_{j_rl_s}(t) \},
\end{equation}
where $P_{jl}(t)$ is the one-grain conditional probability which satisfies eq.
\eqref{master} with the same boundary conditions as the solution
of eq. \eqref{mpmeq}.

As for the single grain, in the multi-grain case the continuous conditional probabilities
$P_{j_1,\dots,j_K;l_1,\dots,l_K}(t)$ are generating functions of the
discrete ones, $G_{j_1,\dots,j_K;l_1,\dots,l_K}(n)$, and we find that
\begin{equation}\label{mggf}
e^{Kt}P_{j_1,\dots,j_K;l_1,\dots,l_K}(t)
=\sum_{n=0}^\infty
G_{j_1,\dots,j_K;l_1,\dots,l_K}(n)\frac{K^nt^n}{n!}.
\end{equation}
The discrete probabilities satisfy the equation
\begin{multline}\label{mgdm}
G_{j_1,\dots,j_K;l_1,\dots,l_K}(n)
=\frac{1}{2K}\sum_{r=1}^K
\{G_{j_1,\dots,j_{r-1},j_r+1,j_{r+1},\dots,j_K;l_1,\dots,l_K}(n-1)
\\
+G_{j_1,\dots,j_{r-1},j_r-1,j_{r+1},\dots,j_K;l_1,\dots,l_K}(n-1)\},
\end{multline}
supplemented by the condition
$G_{j_1,\dots,j_K;l_1,\dots,l_K}(n)=0$, if $j_r=j_{r+1}$, for all $r=1,\dots,K-1$.

\subsection{Infinite lattice}

Let us now consider the asymptotic behaviour for
$t\to\infty$ of the above multi-grain conditional probability. We first
consider the case of an infinite lattice in the horizontal direction when
the one-grain probability is given by eq. \eqref{infty}. Using the integral
representation for the modified Bessel function, we arrive at the expression
\begin{multline}\label{rtinf}
P_{j_1,\dots,j_K;l_1,\dots,l_K}^{(-\infty,\infty)}(t)
=\frac 1{(2\pi)^K }\int_{-\pi }^\pi dx_1\cdots\int_{-\pi }^\pi dx_K\;
e^{-t\sum_{m=1}^K(1-\cos x_m)}
\\ \times
\det_{1\leq r,s\leq K}\left\{e^{i(l_s-j_r)x_r}\right\}.
\end{multline}
Making use of the symmetry of the integrand with respect to permutations of
integration variables $x_1,\dots,x_K$, the determinant in this expression can
be transformed such that
\begin{align}
\det_{1\leq r,s\leq K}\left\{e^{i(l_s-j_r)x_r}\right\}
&\longrightarrow
\det_{1\leq r,s\leq K}\left\{e^{il_sx_r}\right\} \prod_{r=1}^Ke^{-ij_sx_r}
\notag\\
&\longrightarrow
\frac{1}{K!}\det_{1\leq r,s\leq K}\left\{ e^{-ij_sx_r}\right\}
\det_{1\leq r,s\leq K} \left\{e^{il_sx_r}\right\}.
\end{align}
The two determinants above can be represented in terms of Schur functions
(for a survey on Schur functions see  \textit{e.g.} \cite{M-95}):
\begin{align}\label{sch}
s_\lambda (x_1,x_2,\dots,x_K)
:&=\frac{\det_{1\leq s,k\leq K}
(x_s^{\lambda_k+K-k})}{\det_{1\leq s,k\leq K}(x_s^{K-k})}
\notag\\
&=\det_{1\leq s,k\leq K}(x_s^{\lambda _k+K-k})\prod_{1\leq s<k\leq K}(x_s-x_k)^{-1},
\end{align}
where $\lambda =(\lambda _1,\lambda _2,\dots,\lambda_K)$ is a partition
of a non-increasing series of the non-negative integers
$\lambda_1\geq \lambda _2\geq \dots\geq \lambda _K\geq 0$. If we consider
the case $j_r\geq -K$ and $l_r\geq -K$, we find that
\begin{multline}\label{rtinfss}
P_{j_1,\dots,j_K;l_1,\dots,l_K}^{(-\infty,\infty)}(t)
=\frac 1{(2\pi )^KK!}
\int_{-\pi}^\pi dx_1\cdots\int_{-\pi }^\pi dx_K\;
e^{-t\sum_{m=1}^K(1-\cos x_m)}
\\ \times
s_\lambda (e^{ix_1},e^{ix_2},\dots,e^{ix_K})
s_\mu(e^{-ix_1},e^{-ix_2},\dots,e^{-ix_K})
\\ \times
\prod_{1\leq r<s\leq K}|e^{ix_r}-e^{ix_s}|^2,
\end{multline}
where $\lambda_r=j_r-K+r$ and $\mu _r=l_r-K+r$.

As $t\to\infty$
(and $j_s-l_r\ll t$ for all $r,s=1,\dots,K$),
the main contributions to the above integrals come from near the origin of
the integration variables, and in leading order we find that
\begin{multline}\label{Pinfapp}
P_{j_1,\dots,j_K;l_1,\dots,l_K}^{(-\infty,\infty)}(t)\sim
\frac{s_\lambda(1,1,\dots,1)s_\mu (1,1,\dots,1)}{(2\pi )^KK!}
\\ \times
\int_{-\infty }^\infty dx_1\cdots\int_{-\infty }^\infty dx_K\;
e^{-\frac t2\sum_{m=1}^Kx_m^2}\prod_{1\leq r<s\leq K}(x_r-x_s)^2.
\end{multline}
The prefactor of the integral can be computed (see \textit{e.g.} \cite{M-95})
using the well known result
\begin{equation}
s_\lambda (1,1,\dots,1)
=\frac{\prod_{1\leq r<s\leq K}(\lambda _r-r-\lambda _s+s)}{\prod_{m=1}^{K-1}m!},
\end{equation}
while the integral is the Mehta integral of the
gaussian unitary ensemble of random matrices \cite{Me-04}, which can be explicitly
evaluated:
\begin{equation}
\int_{-\infty }^\infty dx_1\cdots\int_{-\infty }^\infty dx_K\;
e^{-\frac 12 t\sum_{m=1}^Kx_m^2}\prod_{1\leq r<s\leq K}(x_r-x_s)^2
=\frac{(2\pi)^{K/2}\prod_{m=1}^Km!}{t^{K^2/2}}.
\end{equation}

We thus find that, as $t\to\infty$, in leading order the multi-grain conditional probability
is given by
\begin{equation}\label{rwa}
P_{j_1,\dots,j_K;l_1,\dots,l_K}^{(-\infty,\infty)}(t)
\sim A_{j_1,\dots,j_K;l_1,\dots,l_K}t^{-\gamma}
\end{equation}
with the scaling exponent
\begin{equation}\label{infgamma}
\gamma =\frac{K^2}2
\end{equation}
and the amplitude
\begin{equation}\label{infamp}
A_{j_1,\dots,j_K;l_1,\dots,l_K}=\frac{\prod_{1\leq s<r\leq
K}(l_r-l_s)(j_r-j_s)}{(2\pi )^{\frac K2}\prod_{m=1}^{K-1}m!}.
\end{equation}

\subsection{Semi-infinite lattice}

Let us consider the conditional probability in the presence of an
absorbing boundary at the origin. As in the one-grain case, let us start with
a finite lattice of $N$ sites in the horizontal direction. Substituting eq. \eqref{hc} into eq.
\eqref{det}, we find that
\begin{multline}
P_{j_1,\dots,j_K;l_1,\dots,l_K}(t)
= \frac{2^K}{(N+1)^K}\sum_{k_1=1}^{N}\dots\sum_{k_K=1}^{N}
e^{-t\sum_{m=1}^KE_{k_m}}
\\ \times
\det_{1\leq r,s \leq K} \left\{ \sin\frac{\pi j_r k_r}{N+1}
\sin \frac{\pi l_s k_r}{N+1}
\right\},
\end{multline}
where $E_k$ is given by eq. \eqref{Ek}.
The multi-grain conditional probability for the semi-infinite
lattice follows from this result by taking the large $N$ limit; the
resulting expression is similar to eq. \eqref{rtinf}, but with a determinant
that now contains sine functions instead of exponential functions:
\begin{multline}\label{sinfk}
P_{j_1,\dots,j_K;l_1,\dots,l_K}^{(0,\infty )}(t)
=\frac 1{\pi^K}\int_{-\pi }^\pi dx_1\cdots\int_{-\pi }^\pi dx_K\;
e^{-t\sum_{m=1}^K(1-\cos x_m)}
\\
\times
\det_{1\leq r,s \leq K} \left\{\sin(j_r x_r)\sin(l_s x_r)\right\}.
\end{multline}
Again, using the symmetry with respect to permutations of the integration
variables $x_1,\dots, x_K$, we can transform the determinant in this
expression such that
\begin{align}
\det_{1\leq r,s\leq K}\left\{\sin(j_rx_r)\sin (l_sx_r)\right\}
&\longrightarrow
\det_{1\leq r,s\leq K}\left\{\sin(l_sx_r)\right\} \prod_{r=1}^K\sin (j_rx_r)
\notag\\
&\longrightarrow
\frac{1}{K!}\det_{1\leq r,s\leq K}\left\{\sin(j_sx_r)\right\}
\det_{1\leq r,s\leq K} \left\{\sin(l_sx_r)\right\}.
\end{align}
Using the character of the irreducible
representation corresponding to a partition $\lambda $ of the
symplectic Lie algebra,
\begin{equation}
sp_\lambda (x_1,x_2,\dots,x_K)
:=\frac{\det_{1\leq j,k\leq K}(x_j^{\lambda _k+K-k+1}-x_j^{-(\lambda _k+K-k+1)})}
{\det_{1\leq j,k\leq K}(x_j^{K-k+1}-x_j^{-(K-k+1)})},
\end{equation}
we can express eq. \eqref{sinfk} in the form
\begin{multline}
P_{j_1,\dots,j_K;l_1,\dots,l_K}^{(0,\infty )}(t)
=\frac 1{\pi^K K!}
\int_{-\pi}^\pi dx_1\cdots\int_{-\pi }^\pi dx_K\;
e^{-t\sum_{m=1}^K(1-\cos x_m)}
\\ \times
\left( \det_{1\leq r,s\leq K}\left\{\sin s x_r\right\} \right)^2
sp_\lambda (e^{ix_1},e^{ix_2},\dots,e^{ix_K})
\\ \times
sp_\mu (e^{ix_1},e^{ix_2},\dots,e^{ix_K}),
\end{multline}
where $\lambda _r=j_r-K+r-1$ and $\mu _r=l_r-K+r-1$. The determinant in the above
integrand can be evaluated using the identity (for a proof, see \cite{KGV-00})
\begin{equation}
\det_{1\leq r,s \leq K} \left\{ \sin sx_r\right\}
=2^{K(K-1)}\prod_{r=1}^K\sin x_r
\prod_{1\leq j<k\leq K}\sin\frac{x_j-x_k}2 \sin\frac{x_j+x_k}2.
\end{equation}
We finally obtain for the conditional probability the expression
\begin{multline} \label{sinfds}
P_{j_1,\dots,j_K;l_1,\dots,l_K}^{(0,\infty )}(t)
=\frac{2^{2K(K-1)}}{\pi^K K!}
\int_{-\pi }^\pi dx_1\cdots\int_{-\pi }^\pi dx_K\;
e^{-t\sum_{m=1}^K(1-\cos x_m)}
\\ \times
\prod_{r=1}^K\sin^2 x_r
\prod_{1\leq j<k\leq K}
\sin^2 \frac{x_j-x_k}2
\sin^2\frac{x_j+x_k}2
\\ \times
sp_\lambda (e^{ix_1},e^{ix_2},\dots,e^{ix_K})
sp_\mu(e^{ix_1},e^{ix_2},\dots,e^{ix_K}).
\end{multline}

In the limit $t\to\infty $ we can approximate the
integrals in the above expression with the integrals
\begin{multline}\label{integral}
\int_{-\infty}^\infty dx_1\cdots
\int_{-\infty}^\infty
dx_K\, e^{-\frac 12 t\sum_{m=1}^K x_m^2}
\prod_{1\leq j<k\leq K}(x_j^2-x_k^2)^2\prod_{j=1}^K x_j^2
\\
=\frac{\prod_{m=1}^K(2m)!}{(2\pi)^{K/2}t^{K(2K+1)/2}}.
\end{multline}
For a proof of eq. \eqref{integral}, see \cite{Me-04}.
We find thereby for the leading asymptotic term of the
generating function
\begin{equation}
P_{j_1,\dots,j_K;l_1,\dots,l_K}^{(0,\infty )}(t)
\sim A_{j_1,\dots,j_K;l_1,\dots,l_K}t^{-\gamma }.
\label{sias}
\end{equation}
Here the scaling exponent is given by
\begin{equation}\label{sinfgamma}
\gamma =\frac{K(2K+1)}2
\end{equation}
and the amplitude is
\begin{equation}\label{sinfamp}
A_{j_1,\dots,j_K;l_1,\dots,l_K}
=\frac{\prod_{m=1}^K(2m)!}{2^{K(K+1)}
\pi^{\frac{3K}2}K!}sp_\lambda (1,\dots,1)sp_\mu (1,\dots,1),
\end{equation}
in which
\begin{equation}
sp_\lambda (1,\dots,1)
=\prod_{1\leq r<s\leq K}\big(j_r^2-j_s^2\big)
\prod_{m=1}^{K-1}
\frac{\left[ 2(K-m)+1\right]!}{m!(K+m)!}.
\end{equation}
A similar expression can be found for $sp_\mu (1,\dots,1)$ with the
$j_r$'s replaced by $l_r$'s.

We thus find that the scaling exponent of the multi-grain sandpile
problem considered here is not equal to the one found previously for the `lock-step' version
of vicious walkers, for which $\gamma=K(K-1)/4$ \cite{F-84,HF-84}. Instead,
our sandpile problem corresponds to the `random-turns' version of vicious
walkers \cite{F-84,Fo-01}. The connection to the `random-turns' version of
vicious walkers is discussed in more detail below.

\section{Connection to vicious walkers}

\subsection{Heisenberg chain}

Before addressing the relation between the above sandpile problem and
the `random-turns' vicious walkers, we first outline
its relation to free fermions. It turns out that it has a straightforward connection
to the Heisenberg XX spin chain that can be mapped, as is well
known, to a free fermion problem by the Jordan-Wigner
transformation.

The Heiseberg XX chain has the Hamiltonian
\begin{equation}\label{haml}
\Hex=-\frac 12\sum_{i,k}\Dex_{ik}\sigma_i^{-}\sigma_k^{+},
\end{equation}
where summation is over all lattice sites, and
\begin{equation}\label{em}
\Dex_{ik}=\delta_{i,k+1}+\delta_{i,k-1}.
\end{equation}
We use the standard notations $\sigma_i^{\pm}, \sigma_i^z$ for Pauli spin operators
that satisfy the commutation relations
\begin{equation} \label{pau}
[\sigma_i^{+},\sigma_k^{-}] =\sigma_i^z\delta _{ik},
\qquad
[\sigma_i^z,\sigma_k^{\pm }] =\pm 2\sigma_i^{\pm}\delta_{ik},
\end{equation}
and have in addition the properties
\begin{equation}\label{nilpotency}
(\sigma^\pm_i)^2=0,\qquad
(\sigma^z_i)^2=1.
\end{equation}
In what follows we use the fact that the ferromagnetic state with
all spins up, $\ket{\Up}=\otimes_i\ket{\uparrow}_i$, which satisfies
$\sigma_k^+\ket{\Up}=0$ for all $k$ and normalized such that $\bracket{\Up}{\Up}=1$,
is annihilated by the Hamiltonian,
\begin{equation}\label{zero}
\hat H\ket{\Up} =0.
\end{equation}
Our aim is to study the `temporal' evolution of states with a finite number
of down spins, which can be constructed by acting with operators
$\sigma_j^-$ on the state $\ket{\Up}$. We thus
consider the matrix elements
\begin{equation}\label{mpcf}
F_{j_1,\dots,j_K;l_1,\dots,l_K}(t)=
\bra{\Up}\sigma_{j_1}^{+}\cdots
\sigma_{j_K}^{+}e^{-t\hat H}\sigma_{l_1}^{-}
\cdots\sigma_{l_K}^{-}\ket{\Up}.
\end{equation}
Parameter $t$ will play the role of `time' in the context of the
sandpile model.

Before proceeding with the general case, let us first consider
the case $K=1$, \textit{i.e.} the `temporal' evolution of a single reversed spin.
Differentiating the function
$F_{jl}(t)= \bra{\Up}\sigma_j^{+}e^{-t\hat H}\sigma_l^{-}\ket{\Up}$
with respect to $t$ and using the commutation relation
\begin{equation}\label{cr}
[\sigma_j^{+},\Hex] =-\frac 12\sum_k\Dex_{jk}
\sigma_j^z\sigma_k^{+}=-\frac 12\sigma_j^z(\sigma_{j-1}^{+}+\sigma_{j+1}^{+})
\end{equation}
together with the property $\bra{\Up}\sigma_j^z=\bra{\Up}$, we find that
\begin{equation}
\frac d{dt}F_{jl}(t)= -\bra{\Up} \sigma _j^{+}
\hat He^{-t\hat H}\sigma _l^{-}\ket{\Up}
=\frac 12 \bra{\Up}
(\sigma_{j-1}^{+}+\sigma_{j+1}^{+})e^{-t\hat H}\sigma_l^{-}
\ket{\Up}.
\end{equation}
Hence the correlation function eq. \eqref{mpcf} for $K=1$ satisfies the equation
\begin{equation}\label{eseq}
\frac d{dt}F_{jl}(t)=\frac{1}{2}\left(
F_{j+1l}(t)+F_{j-1l}(t)\right).
\end{equation}
Similarly, by commuting $\Hex$ with $\sigma_l^-$, a difference equation similar
to eq. \eqref{eseq} follows, but for subscript $j$ with fixed subscript $l$. Both
equations are subject to the initial condition $F_{jl}(0)=\delta_{jl}$, and
to boundary conditions that depend on the type of the lattice:
for the semi-infinite lattice $F_{jl}=0$ for $j,l=0$,
while for a finite lattice $F_{jl}=0$ for $j,l=0$ and $j,l=N+1$.

As a result, comparing eqs. \eqref{eseq} and \eqref{master} together with their
initial and boundary conditions, we find that the one-spin
correlation function of the Heisenberg XX chain is equal, modulo a trivial
factor, to the one-grain conditional probability of the sandpile model, \textit{i.e.}
$P_{jl}(t)=e^{-t} F_{jl}(t)$.

Let us now consider the case of general $K$.
Differentiating eq. \eqref{mpcf} with respect to $t$,
taking into account the differential property of the commutation relations,
\begin{equation} 
[\sigma _{j_1}^{+}\sigma _{j_2}^{+}\cdots
\sigma_{j_K}^{+},\Hex]
=\sum_{k=1}^K\sigma _{j_1}^{+}\cdots\sigma_{j_{k-1}}^{+}
[\sigma _{j_k}^{+},\Hex]\sigma_{j_{k+1}}^{+}\cdots\sigma _{j_K}^{+},
\end{equation}
and applying the commutation relation  eq. \eqref{cr}, we find that
\begin{multline}\label{phem}
\frac{d}{dt}
F_{j_1,\dots,j_K;l_1,\dots,l_K}(t)
=\frac 12\sum_{r=1}^K
\Big(F_{j_1,\dots,j_{r-1},j_r+1,j_{r+1},\dots,j_K;l_1,\dots,l_K}(t)
\\
+F_{j_1,\dots,j_{r-1},j_r-1,j_{r+1},\dots,j_K;l_1,\dots,l_K}(t)
\Big).
\end{multline}
A similar equation can be found with respect to subscripts $l_r$
with the $j_r$'s kept fixed. The initial condition is
$F_{j_1,\dots,j_K;l_1,\dots,l_K}(0)=\delta_{j_1l_1}\cdots\delta_{j_Kl_K}$.
The correlation function also satisfies
the conditions $F_{j_1,\dots,j_K;l_1,\dots,l_K}(t)=0$
if $l_r=l_s$ or $j_r=j_s$ ($r,s=1,\dots,K$), which follow from the
nilpotency of the Pauli spin operators, eq. \eqref{nilpotency}.

It is evident that the differential equation eq. \eqref{phem} which the
correlation function eq. \eqref{mpcf} satisfies, coincides with
eq. \eqref{mpmeq} for the multi-grain probability up to a trivial
`diagonal' term. It is also easy to verify that the solution of
eq. \eqref{phem} can be expressed in a determinant form,
\begin{equation}\label{detcf}
F_{j_1,\dots,j_K;l_1,\dots,l_K}(t)=
\det_{1\leq r,s \leq K} \left\{ F_{j_rl_s}(t)\right\},
\end{equation}
where $F_{jl}(t)$ are the one-particle correlation functions satisfying
eq. \eqref{eseq}. We thus find that
\begin{equation}\label{P=F}
P_{j_1,\dots,j_K;l_1,\dots,l_K}(t)=e^{-Kt}
F_{j_1,\dots,j_K;l_1,\dots,l_K}(t).
\end{equation}
Hence, the multi-grain probabilities in our sandpile problem are also
multi-spin correlation functions in the Heisenberg $XX$ spin chain.

\subsection{Quantum trajectories}

We are now in the position to establish an explicit relation of our sandpile
model with random walks. To this end we exploit the known connection of the
Heisenberg XX spin chain with the random-turns vicious walk
model \cite{B-06,B-07,BM-09}. Our starting point, which
follows from eqs. \eqref{P=F} and \eqref{mggf}, is that the discrete
multi-grain probability can be expressed as a matrix element:
\begin{equation}\label{cprepmg}
G_{j_1,\dots,j_K;l_1,\dots,l_K}(n)
=\frac{1}{K^n}
\bra{\Up}\sigma_{j_1}^{+}\cdots
\sigma_{j_K}^{+}(-\Hex)^n\sigma_{l_1}^{-}
\dots\sigma_{l_K}^{-}\ket{\Up}.
\end{equation}
By evaluating this matrix element, it can be shown that the
discrete probability of the sandpile model is equal, modulo a simple
factor, to the number of paths of random-turns vicious walkers subject to
the same boundary conditions.

We start again with the single-grain case in which
the discrete probability $G_{jl}(n)$ is the one-spin matrix element
$\bra{\Up} \sigma_j^{+}(-\Hex)^n\sigma_l^{-}\ket{\Up}$. Taking into account eq.
\eqref{zero}, we can write
\begin{equation}\label{lp}
\Hex\sigma_l^{-} \ket{\Up}
=\big[\Hex,\sigma_l^{-}\big]\ket{\Up}
=-\frac 12\sum_{k_1} \Dex_{k_1l}\;\sigma_{k_1}^{-}\ket{\Up},
\end{equation}
and repeating the procedure for the $n$th power, we find that
\begin{equation}\label{single}
\Hex^n\sigma_l^{-} \ket{\Up}=\frac {(-1)^n}{2^n}\sum_{k_1,\dots,k_n}
\Dex_{k_nk_{n-1}}\cdots \Dex_{k_2k_1}\Dex_{k_1l}\; \sigma_{k_n}^{-}\ket{\Up}.
\end{equation}
Multiplying this expression from the left by
$\bra{\Up}\sigma_j^{+}$ and using the orthogonality of the spin states,
$\bra{\Up}\sigma_j^{+}\sigma_l^{-}\ket{\Up}=\delta_{jl}$, we find that
\begin{equation}\label{onespin}
\bra{\Up} \sigma_j^{+}(-\Hex)^n\sigma_l^{-}\ket{\Up}=\frac{1}{2^n}
\sum_{k_1,\dots,k_{n-1}}\Dex_{jk_{n-1}}\cdots\Dex_{k_2k_1} \Dex_{k_1l}.
\end{equation}
The sum in eq. \eqref{onespin} can be
interpreted as one over all
possible quantum trajectories (lattice paths) of $n$ time steps of a particle
(corresponding to the down spin state) from site $l$ to site $j$ subject
to the boundary conditions. In this interpretation $\Dex$ appears as the
transfer matrix. From eq. \eqref{em} it follows that we deal with a random
walk on a lattice. Denoting by $\Path_n(l\mapsto j)$ the number of all
admissible paths of $n$ steps from site $l$ to site $j$, we have
\begin{equation}\label{gpi}
G_{jl}(n)=\frac 1{2^n}\,\Path_n(l\mapsto j).
\end{equation}

Let us now consider the general case. Acting with the Hamiltonian $\Hex$
on the state $\sigma_{l_1}^{-}\sigma_{l_2}^{-}\cdots\sigma_{l_K}^{-}\ket{\Up}$,
for which we assume that $l_1>l_2>\dots>l_K$, we find that
\begin{align} 
\Hex\sigma_{l_1}^{-}\sigma_{l_2}^{-}\cdots\sigma_{l_K}^{-}\ket{\Up}
&=\sum_{r=1}^K\sigma_{l_1}^{-}\cdots\sigma_{l_{r-1}}^{-}
\left[\Hex,\sigma _{l_r}^{-}\right]\sigma_{l_{r+1}}^{-}
\cdots\sigma_{l_K}^{-}\ket{\Up}
\notag\\
&=-\frac{1}{2}\sum_{r=1}^K\sum_{m}
\Dex_{m l_r}\sigma_{l_1}^{-}
\cdots\sigma_{l_{r-1}}^{-}\sigma_{m}^{-}\sigma_{l_{r+1}}^{-}
\cdots\sigma_{l_K}^{-}\ket{\Up},
\notag\\
&=-\frac{1}{2}\sum_{m_1,\dots,m_K}
T_{m_1,\dots,m_K;l_1,\dots,l_K}
\sigma_{m_1}^{-}
\cdots\sigma_{m_K}^{-}\ket{\Up},
\end{align}
where
\begin{equation}\label{T}
T_{m_1,\dots,m_K;l_1,\dots,l_K}
=\sum_{r=1}^K \delta_{m_1l_1}\cdots\delta_{m_{r-1}l_{r-1}}
\Dex_{m_r l_r}\delta_{m_{r+1}l_{r+1}}\cdots\delta_{m_K l_K}
\end{equation}
for $m_1>m_2>\dots>m_K$, and $T_{m_1,\dots,m_K;l_1,\dots,l_K}=0$ for
$m_r=m_{r+1}$ ($r=1,\dots,K-1$). Interpreting $T$ as a transfer matrix,
we find for the multi-spin matrix element an expression in terms of
a matrix element of the $n$th power of this transfer matrix,
\begin{equation}
\bra{\Up}\sigma _{j_1}^{+}\dots\sigma _{j_K}^{+}
(-\Hex)^n\sigma_{l_1}^{-}\cdots\sigma_{l_K}^{-}\ket{\Up}
=\frac{1}{2^n} (T^n)_{j_1,\dots,j_K;l_1,\dots,l_K},
\end{equation}
which generalizes eq. \eqref{single} to the case of $K$ random walkers.

The presence of just single factor $\Dex$ in each term of the transfer
matrix eq. \eqref{T} implies that, at each time step, only a single
walker moves out of the total $K$. Thus the above matrix element of the
$n$th power of $T$ gives the number of all lattice paths of $n$ steps
made by $K$ \emph{random-turns} vicious walkers. We recall that, in the
random-turns vicious walkers model, at
each time step only a single randomly chosen walker moves
one step to the `left' or one step to the `right' with
the constraint that two walkers cannot occupy the same site.
These random-turns vicious walkers are different from the more
common lock-step vicious walkers \textit{all} of which,
at each time step, must move left or right with the same constraint
that two walkers cannot occupy the same site \cite{F-84}.

Denoting by $\Path_n(l_1,\dots,l_K\mapsto j_1,\dots,j_K)$
the number of all admissible configurations in which the $K$
walkers are initially located on the lattice
sites $l_1>l_2>\dots>l_K$, and have after $n$ steps arrived at the
positions $j_1>j_2>\dots>j_K$,  we find the result
\begin{equation}\label{gpimg}
G_{j_1,\dots,j_K;l_1,\dots,l_K}(n)=
\frac 1{(2K)^n}\Path_n(l_1,\dots,l_K\stackrel{\text{RT}}{\mapsto} j_1,\dots,j_K),
\end{equation}
where RT stands for random-turns vicious walks.

\subsection{Large $n$ limit}

Having established a connection of our sandpile model
with the random-turns vicious walkers, it is natural to consider
the large $n$ limit of discrete conditional probabilities.
This is useful for a direct comparison with the random-turns walkers, for
both an infinite and a semi-infinite lattice in the horizontal direction.

For definiteness we consider here an infinite lattice in the horizontal
direction; a semi-infinite lattice can be considered similarly, and below
we outline the results for both cases. Using the relation
between the continuous and discrete conditional probabilities,
see eq. \eqref{mggf}, we find from eq. \eqref{rtinf} in the case of an
infinite lattice the representation
\begin{multline}\label{ginf}
G_{j_1,\dots,j_K;l_1,\dots,l_K}^{(-\infty,\infty)}(n)
=\frac 1{(2\pi)^K K^n}\int_{-\pi }^\pi dx_1\cdots\int_{-\pi }^\pi dx_K\;
\left(\sum_{m=1}^K\cos x_m\right)^n
\\ \times
\det_{1\leq r,s\leq K}\left\{e^{i(l_s-j_r)x_r}\right\}.
\end{multline}
We are interested here in the large $n$ limit with the $l_r$'s and $j_r$'s
kept fixed. In order to apply the standard saddle-point approximation, we
express the first factor of the integrand in the above equation in the
form $\exp\left\{n\log  \left(\sum_m\cos x_m\right)\right\}$, and
obtain thereby the following system of saddle-point equations:
\begin{equation}\label{sp}
\frac{\sin x_r}{\sum_{m=1}^K\cos x_m}=0, \qquad r=1,\dots,K.
\end{equation}
It is evident that the solutions to this system of equations satisfy
$\sin x_r=0$($r=1,\dots,K$) with the restriction that
$\sum_m \cos x_m\ne 0$. Requiring that the matrix of second derivatives
\begin{multline}
\frac{\partial^2}{\partial x_rx_s}\log\left(\sum_{m=1}^K\cos x_m\right)
=-\frac{\cos x_r}{\sum_{m=1}^K\cos x_m}\,\delta_{rs}
-\frac{\sin x_r\sin x_s}{\left(\sum_{m=1}^K\cos x_m\right)^2}
\end{multline}
is a negatively definite matrix for the solution of eq. \eqref{sp}, we
find that the steepest descent corresponds to the solution for which
$\cos x_r=1$ ($r=1,\dots,K$) \textit{i.e.} the main contribution to the
integrals in eq. \eqref{ginf} comes from near the points
$x_r=0$ ($r=1,\dots,K$), similarly to the case of continuous conditional
probability considered in sect. 3.2.

Therefore, replacing the first factor of the integrand in eq. \eqref{ginf}
by its approximation near the origin of the integration variables, \textit{i.e.}
\begin{equation}\label{}
\left(\sum_{m=1}^K\cos x_m\right)^n\propto K^n \exp\left\{-\frac{n}{2K}\sum_{m=1}^K x_m^2\right\},
\end{equation}
and transforming the second factor of the integrand as in sect. 3.2, we
find that as $n\to\infty$ the leading order form of the discrete
multi-grain probability in the case of an infinite lattice can be
expressed as
\begin{multline}
G_{j_1,\dots,j_K;l_1,\dots,l_K}^{(-\infty,\infty)}(n)\sim
\frac{s_\lambda(1,1,\dots,1)s_\mu (1,1,\dots,1)}{(2\pi )^KK!}
\\ \times
\int_{-\infty }^\infty dx_1\cdots\int_{-\infty }^\infty dx_K\;
e^{-(n/2K)\sum_{m=1}^Kx_m^2}
\prod_{1\leq r<s\leq K}(x_r-x_s)^2.
\end{multline}
Clearly this expression is the same as eq. \eqref{Pinfapp} with
$t$ replaced by $n/K$. It is easy to check that a similar result
is valid for a semi-infinite lattice, now using the procedure of
sect. 3.3. Hence, the leading terms of the large
$n$ limits of discrete multi-grain probabilities can be obtained
from those of the large $t$ limits of the continuous probabilities
simply by the replacement $t\mapsto n/K$.

We have thus shown that, as $n\to \infty$ for fixed $l_r$'s
and $j_r$'s, the discrete multi-grain conditional probabilities
scale as
\begin{equation}\label{}
G_{j_1,\dots,j_K;l_1,\dots,l_K}(n)
\sim B_{j_1,\dots,j_K;l_1,\dots,l_K}\,n^{-\gamma}
\end{equation}
with
\begin{equation}\label{}
B_{j_1,\dots,j_K;l_1,\dots,l_K}=K^{\gamma} A_{j_1,\dots,j_K;l_1,\dots,l_K},
\end{equation}
where the exponent $\gamma$ and amplitude
$A_{j_1,\dots,j_K;l_1,\dots,l_K}$ are given by eqs. \eqref{infgamma}
and \eqref{infamp}, respectively, in the case of an infinite lattice,
and by eqs. \eqref{sinfgamma} and \eqref{sinfamp}, respectively, in the
case of a semi-infinite lattice. The discrete forms of the conditional
probabilities scale exactly as the continuous ones, as they should. These
results, in view of eq. \eqref{gpimg}, are also in agreement with the
known scaling properties of the random-turns vicious walkers \cite{Fo-91}.


\section{Conclusion}

In conclusion, we showed that the probability of multiple grains in a two
dimensional directed sandpile to preserve their order during an avalanche,
displays non-trivial scaling properties that are similar to those of the
`ran\-dom-turns' version of vicious walkers. The conditional probability
in the case when the downhill direction of the lattice was transformed into
a continuous variable (`time'), which we found as the solution to a master
equation, was found to be the generating function of directed lattice paths
in the original lattice. This continuous time conditional probability then
provided the connection with the Heisenberg XX spin chain as it was found to
be the same as the corresponding many-spin correlation function of the
Heisenberg chain. Notice that connection with spin-$1/2$ variables was
based on having a directional lattice with the chosen initial condition.
After a toppling on `row' $n$ there were only sites with a critical or
subcritical (critical minus one) number of grains on row $n+1$, i.e. there were
effectively only two possible states per site. For other situations connection
would possibly be to spin systems of higher spin. Connection with the
Heisenberg chain also established a relation between sandpile models and
free fermions. Similarly to the sandpile problem, the spin correlation
function was shown to be the generating function of `vicious' quantum
trajectories of spins.

\section*{Acknowledgements}
This research was partially supported by the Russian Foundation for Basic
Research, grant 10-01-00600, the Russian Academy of Sciences programme
Mathematical Methods in Nonlinear Dynamics and the University of
Jyv\"{a}skyl\"{a}. A.G.P. was supported by the
Alexander von Humboldt Foundation research fellowship.

\bibliography{sandpile_bib}
\end{document}